\documentclass[twocolumn]{aastex63}
\usepackage{xspace}
\newcommand{\hi}{{\sc H\,i}\xspace}
\newcommand{\snyr}{SN yr$^{-1}$\xspace}
\newcommand{\pfin}{$p_{\rm fin}$\xspace}
\newcommand{\snyrperA}{SN yr$^{-1}$ kpc$^{-2}$\xspace}
\newcommand{\sigrat}{$\sigma^{min}_p/\sigma_{obs}$\xspace}

\usepackage{amsmath}
\usepackage{gensymb}
\usepackage{subfigure}
\usepackage{amssymb}

\shorttitle{Feedback from observations}
\shortauthors{Sarbadhicary et. al}

\begin{document}

\title{Testing the Momentum-driven Supernova Feedback Paradigm in M31}

\correspondingauthor{Sumit K. Sarbadhicary}
\email{sarbadhicary.1@osu.edu}

\author[0000-0002-0786-7307]{Sumit K. Sarbadhicary}
\affil{Department of Physics and Astronomy, Michigan State University, East Lansing, MI 48824, USA}
\affil{Center for Cosmology and AstroParticle Physics (CCAPP), Ohio State University, Columbus, OH 43210, USA}

\author{Davide Martizzi}
\affiliation{DARK, Niels Bohr Institute, University of Copenhagen, Blegdamsvej 17, 2100 Copenhagen, Denmark}

\author{Enrico Ramirez-Ruiz}
\affiliation{Department of Astronomy and Astrophysics, University of California, Santa Cruz, CA 95064, USA}
\affiliation{DARK, Niels Bohr Institute, University of Copenhagen, Blegdamsvej 17, 2100 Copenhagen, Denmark}

\author{Eric Koch}
\affiliation{University of Alberta, Department of Physics, 4-183 CCIS, Edmonton AB T6G 2E1, Canada}
\affiliation{Center for Astrophysics $\mid$ Harvard \& Smithsonian, 60 Garden St., Cambridge, MA 02138, USA}

\author{Katie Auchettl}
\affiliation{School of Physics, 
 The University of Melbourne, 
 Parkville, VIC 3010, Australia}
\affiliation{ARC Centre of Excellence 
 for All Sky Astrophysics in 3 Dimensions 
 (ASTRO 3D)}
\affiliation{Department of Astronomy and Astrophysics, University of California, Santa Cruz, CA 95064, USA}

\author{Carles Badenes}
\affiliation{Department of Physics and Astronomy and Pittsburgh Particle Physics, Astrophysics and Cosmology Center (PITT PACC),
University of Pittsburgh, 3941 O’Hara Street, Pittsburgh, PA
15260, USA}

\author{Laura Chomiuk}
\affil{Department of Physics and Astronomy, Michigan State University, East Lansing, MI 48824, USA}



\begin{abstract}
Momentum feedback from isolated supernova remnants (SNRs) have been increasingly recognized by modern cosmological simulations as a resolution-independent means to implement the effects of feedback in galaxies, such as turbulence and winds. However, the integrated momentum yield from SNRs is uncertain due to the effects of SN clustering and interstellar medium (ISM) inhomogeneities. In this paper, we use spatially-resolved observations of the prominent 10-kpc star-forming ring of M31 to test models of mass-weighted ISM turbulence driven by momentum feedback from isolated, non-overlapping SNRs. We use a detailed stellar-age distribution (SAD) map from the Panchromatic Hubble Andromeda Treasury (PHAT) survey, observationally-constrained SN delay-time distributions, and maps of the atomic and molecular hydrogen to estimate the mass-weighted velocity dispersion using the Martizzi et al. ISM turbulence model. Our estimates are within a factor of 2 of the observed mass-weighted velocity dispersion in most of the ring, but exceed observations at densities $\lesssim 0.2$ cm$^{-3}$ and SN rates $>2.1\times 10^{-4}$ SN yr$^{-1}$ kpc$^{-2}$, even after accounting for plausible variations in stellar-age distribution models and ISM scale height assumptions. We conclude that at high SN rates the momentum deposited is most likely suppressed by the non-linear effects of SN clustering, while at low densities, SNRs reach pressure equilibrium before the cooling phase. These corrections should be introduced in models of momentum-driven feedback and ISM turbulence.
%
\end{abstract}

\keywords{editorials, notices --- 
miscellaneous --- catalogs --- surveys}


\section{Introduction} \label{sec:intro}
Supernova (SN) feedback plays a critical role in galaxy formation by regulating the phase structure of the interstellar medium (ISM; \citealt{MO77, MacLow2004, Joung2006}), launching galactic winds \citep{Strickland2009, Heckman2017, Zhang2018}, accelerating cosmic rays \citep{Drury1994, Socrates2008, Caprioli2011, Girichidis2016}, and enriching the intergalactic and circumgalactic medium with metals \citep{Andrews2017, Weinberg2017, Telford2018}. Through a combination of these phenomena, feedback regulates the global star-formation efficiency of galaxies \citep{Ostriker2011, Hopkins2012}. Simulations imply that, without feedback, galaxies would rapidly convert cold gas into stars, resulting in up to a factor of 100 overproduction of stars compared to what is observed \citep{Navarro1991, Hopkins2011}. 

Unfortunately, current state-of-the-art cosmological simulations that study the evolution of galaxy population over cosmic time cannot resolve the spatial scales on which supernova remnants (SNRs) interact with the ISM. Even modern `zoom-in' simulation of isolated galaxies can only marginally resolve SNRs \citep{Hopkins2014, Hopkins2018}, and properly resolved SNRs can only be obtained in simulations of smaller regions of the ISM disk \citep{Gatto2017, Kim2017,2020ApJ...896...66K}. This limitation motivated development of subgrid models of SN feedback at the resolution limit of simulations.
Initial efforts to implement SN feedback in the form of thermal energy deposition were ineffective due to efficient radiative cooling 
in high-density star-forming regions \citep{Katz1992}. The quest to limit plasma cooling and runaway star-formation spawned a variety of subgrid models which employed techniques like delayed gas-cooling \citep{Stinson2006, Governato2007, Governato2010}, stochastic thermal feedback \citep{Dalla2012}, an effective equation of state for a pressure-supported multi-phase ISM, with hydrodynamically decoupled wind particles \citep{Springel2000, Springel2003, Oppenheimer2006, Vogelsberger2014} These techniques ranged from being unphysical in nature to being inaccurate in the details of the SN-ISM coupling \citep{Martizzi2015, Rosdahl2017, Smith2018}.

More recent cosmological simulations \citep[e.g., FIRE,][]{Hopkins2018, Hopkins2018b} have explored subgrid models that deposit momentum, which unlike thermal energy, cannot be radiated away before impacting ambient gas \citep{Murray2005, Socrates2008, Agertz2013}. During the Sedov-Taylor phase of SNRs, the blast wave increases its momentum yield by a factor of 10--30 as it sweeps up ambient ISM. Later, it transitions into a cold, dense, momentum-conserving shell that ultimately merges with the ISM \citep{Chevalier1974, Cioffi1998, Thornton1998, Martizzi2015, Kim2015, 2020ApJ...896...66K}. This momentum budget per SN has been quantified by several realistic models of the ISM \citep[e.g.,][]{Martizzi2015, Kim2015, Li2015, Walch2015,2020ApJ...896...66K}. It has been shown to effectively drive turbulence and winds, and reproduce key features of galaxies such as the Kennicutt-Schmidt relation and galactic morphologies \citep{Martizzi2016, Smith2018, Hopkins2018}. 
More recent studies however have shown that the momentum deposition per SN depends sensitively on effects like SN clustering \citep{Sharma2014, Gentry2017}, entrainment of cold clouds \citep{Pittard2019}, the abundance pattern of the ISM \citep{2020ApJ...896...66K}, thermal conduction \citep{Badry2019}, enhanced cooling due to fluid-instability-driven mixing across the contact discontinuity \citep{Gentry2019}, the SN delay-time distribution (DTD) model \citep{Gentry2017, Keller2020}, and pre-SN feedback via winds, photoionization and radiation pressure \citep{Fierlinger2016, Smith2020}

Observations that are specifically sensitive to SN feedback can identify a reliable subgrid model. Generally, cosmological simulations calibrate subgrid models to reproduce bulk properties of the galaxy population such as the stellar mass function and stellar mass to halo mass relation, but this necessarily limits the predictive power of the simulations \citep{Schaye2015}. Extragalactic multi-wavelength have served as useful references for setting subgrid model components such as SN rates \citep[e.g.][]{Mannucci2006}, the efficiency of SN energy driving ISM turbulence \citep[e.g.][]{Tamburro2009, Stilp2013} and mass-loading in supersonic winds \citep[e.g.][]{Martin1999, Veilleux2005, Strickland2009}. However, the main source of uncertainty in modern subgrid models stem from a poor understanding of the SN-ISM interaction physics that originates on scales of 10-100 pc, which is beyond the reach for most distant surveys. In this respect, the resolved environments of Local Group galaxies provide detailed information available on stellar populations, ISM distribution and kinematics, and SNRs at the highest affordable spatial resolution. They are the ideal testing grounds for SN feedback models.

In this work, we test models of mass-weighted ISM turbulence predicted by SN momentum feedback models against observations of turbulence in M31, with a focus on the long-lived, prominent 10-kpc star-forming ring \citep{Lewis2015, Williams2015}. The proximity of M31 pushes the frontier of  turbulence studies to $<100$ pc, where the effects of feedback are spatially-resolved, complete with detailed maps of the atomic ISM distribution \citep[e.g.][]{Braun1991,Nieten2006,Braun2009}, and spatially-resolved stellar age distribution (SAD) measurements with sensitivity down to masses $\approx 1.5$ M$_{\odot}$ obtained by the Panchromatic Hubble Andromeda Treasury survey \citep[PHAT;][]{Dalcanton2012, Williams2017}. We can use these SADs to estimate SN rates by taking into account currently known constraints on the efficiencies of the different progenitor channels of core-collapse and Type Ia SNe, expressed in the form of SN delay-time distributions (DTDs) \citep{Maoz2014, Zapartas17, Eldridge2017}. The SADs of the older stellar populations allow us to quantify the Type Ia SN rate as a function of location, which is important for a `green-valley' galaxy like M31 \citep{Mutch2011, Davidge2012}, and lacks correlation with conventional star-formation rate tracers.

This paper is organized as follows. In Section \ref{sec:model} we describe our analytical momentum-driven ISM turbulence model, and how we use stellar population and ISM data to constrain ISM densities, SN rates and velocity dispersion in the M31 ring. Section \ref{sec:results} describes the results of our analysis and checks on potential systematics, and in Section \ref{sec:disc}, we discuss the implications of these results on the assumed subgrid models of feedback used in cosmological simulations.

\section{Modeling ISM Velocity Dispersion in M31} \label{sec:model}
Here, we compare the observed non-thermal velocity dispersion in M31's neutral (\hi) and cold ISM 
with the predicted turbulent velocity dispersion from the SN momentum-driven ISM turbulence model of \cite{Martizzi2016}.
Our calculations are supplemented by measured SN rates from the SAD of the PHAT survey and known forms of the SN DTD.
We describe these efforts below. For all measurements, we assume that the distance to M31 is 785 kpc \citep{McConnachie2005}, and 1\arcsec = 3.78 pc at the distance of M31.

\begin{figure} 
\includegraphics[width=\columnwidth]{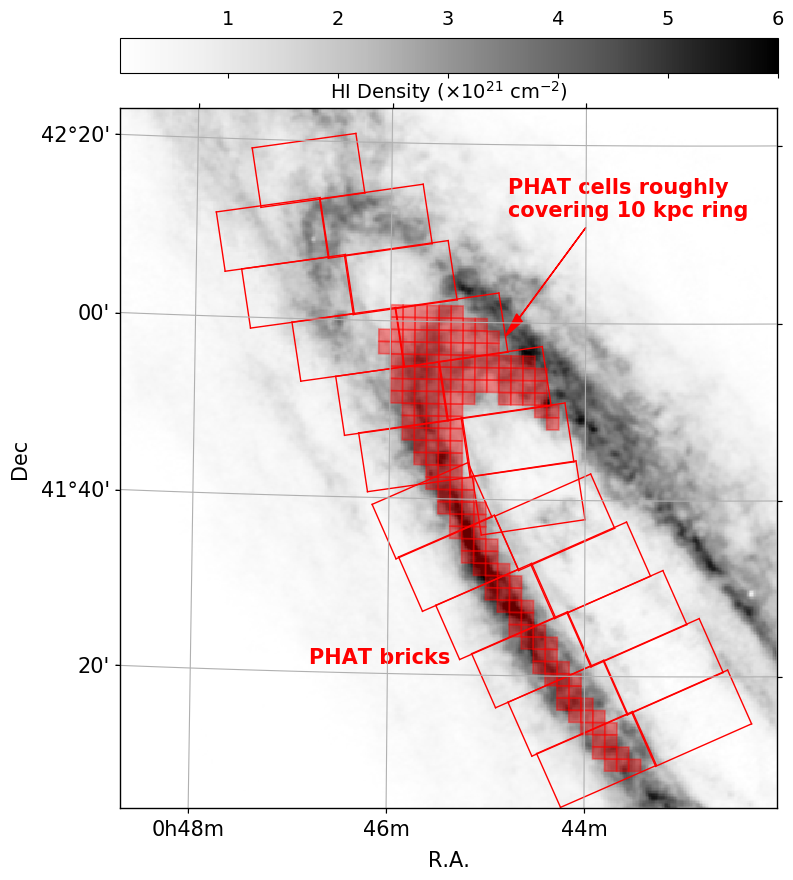}
\caption{Map of \hi column density derived from 21 cm observations by \cite{Braun2009}. Overlaid is the footprint of the PHAT survey, with observation bricks outlined as large red rectangles \citep{Dalcanton2012}. Stellar age distributions are measured in 83$^{\prime\prime}$ cells within the PHAT area \citep{Williams2017}. The shaded red squares show the locations of cells located between deprojected radii of 10--13 kpc from M31's center; they roughly cover the main star-forming ring of M31 (see Section \ref{subsec:ring}). We will compare our model (Section \ref{sec:model}) with the observed velocity dispersion in this ring.}
\label{fig:maps}
\end{figure}

We restrict our analysis to the 10 kpc star-forming ring of M31 (Figure \ref{fig:maps}). We expect the main source of turbulence here to be star-formation, which we are mainly interested in testing, as opposed to other sources of turbulence observed in galaxies such as galactic spiral arms and magnetorotational instabilities \citep{Tamburro2009,Koch2018,utomo2019}. Both atomic and molecular gas are most abundantly located and detected at high signal-to-noise along the ring \citep{Braun2009, Nieten2006}. Additionally, since the gas scale height in pressure-supported star-forming disks can vary with radius \citep{utomo2019}, staying within the ring helps justify the use of a constant scale height in Eq.\ \eqref{eq:nh1} and \eqref{eq:nh2}. We do however assess the impact of variable scale heights later on in Section \ref{sec:modelvsobsveldisp}. 

In the following sub-sections, we describe our methodology for modeling the ISM velocity dispersion using momentum-driven turbulence from SNe, and the observations we use for comparison.

\subsection{Momentum Injection by Supernovae} \label{sec:predsigma}

Following \cite{Martizzi2015} and \cite{Martizzi2016} -- hereafter, M15 and M16 respectively -- we assume that the non-thermal velocity dispersion in \hi is a result of SNR momentum-driven turbulence driven on spatial scales comparable to the radius at which the SNR merges with the ISM, i.e. the shock velocity becomes of the order of velocity dispersion in the ISM. M15 quantified the final momentum (\pfin) driven by an isolated SNR well past its shell formation stage in a turbulent ISM as

\begin{equation} \label{eq:pfin_nh}
\frac{p_{\rm fin}}{m_*} = 1110\ \mathrm{km/s} \left(\frac{Z}{Z_{\odot}}\right)^{-0.114} \left(\frac{n_h}{100\ \mathrm{cm^{-3}}}\right)^{-0.19}
\end{equation}

where \pfin/$m_*$ is the momentum deposited per mass of stellar population (we set $m_* = 100$ M$_{\odot}$ per M15), $Z$ is the metallicity and $n_h$ is the ISM density\footnote{The reader is refer to \citet{2020ApJ...896...66K} for revised prescriptions at low $Z$}. We note that this form of \pfin is similar to other independent high-resolution studies of momentum deposition by SNRs in an inhomogenous ISM \citep[e.g.][]{Kim2015}.

M16 used this subgrid model of momentum feedback to simulate the SN-driven ISM at 2--4 pc spatial resolution, and showed that the resulting velocity dispersion in a steady-state Milky Way-like ISM can be described by an analytical equation where the energy injection rate of SN momentum-driven turbulence is balanced by its corresponding rate of decay. The resulting mass-weighted velocity dispersion ($\sigma_p$) is given by the Eq 22. in M15, which we repeat here for convenience, 

\begin{equation}
\label{eq:sigma}
 \sigma_{p} = \dfrac{3}{4\pi} \left(\dfrac{32\pi^2}{9}\right)^{3/7} \left(\dfrac{p_{\rm fin}}{\rho}\right)^{4/7} \left(f\, \dot{n}_{SN}\right)^{3/7}
 \end{equation}
where $\rho$ is the density of gas, $\dot{n}_{SN}$ is the SN rate per unit volume, and $f$ is a factor that accounts for momentum cancellation when multiple blast waves interact. We set $f=1$ for our fiducial runs, then revisit the issue of $f$ in Section \ref{sec:disc}. We will use this predicted $\sigma_p$ in different regions of M31's ring, as a function of the measured $\rho$ and $\dot{n_{SN}}$, for comparison with observations in the subsequent sections.
    
\begin{figure}
    \centering
    \includegraphics[width=\columnwidth]{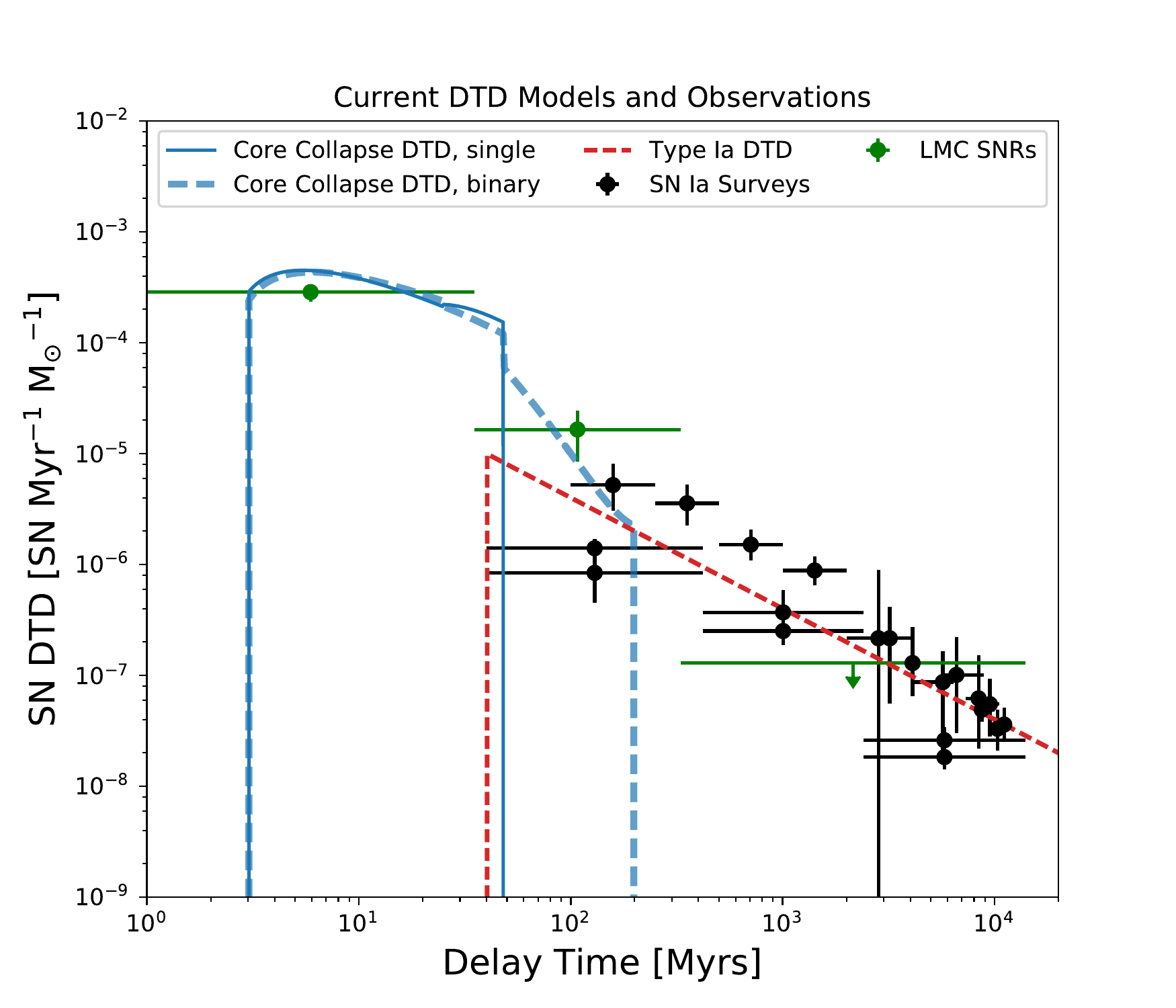}
    \caption{Form of the SN DTD for core-collapse and Type Ia SNe. The solid blue line shows the DTD for core-collapse \citep{Zapartas17} assuming single stellar evolution, while the dashed line model shows the extended delay-time tail due to binary evolution. Dashed red line shows the form of the Type Ia DTD used in this paper, in comparison with observations from SN Ia surveys \citep{Totani2008,Maoz2010a, Maoz2011, Maoz2012b} and SNR surveys \citep{Maoz2010}. See Section \ref{sec:ratesmodel} for details.}
    \label{fig:dtd}
\end{figure}
\subsection{Measurement of SN rate density ($\dot{n}_{SN}$)} \label{sec:ratesmodel}

We set the SN rate per unit volume, $\dot{n}_{SN}$ using the detailed SAD maps from the PHAT survey and the known properties of SN DTD.  We use the \cite{Williams2017} SAD map of the northern third of the disk of M31, spanning a total area of about 0.8 deg$^2$ (Figure \ref{fig:maps}). An SAD is measured in each of 826 spatial cells, each 83$^{\prime\prime}$ wide. Each cell contains the stellar mass formed per look-back time ($M_{ij}$, where $i$ is the cell and $j$ is the age bin), estimated by comparing resolved color-magnitude diagrams of the stars in the region with stellar isochrone models. 

We then convert these SAD maps into maps of the SN rate per cell using observationally constrained DTDs. The DTD is defined as the SN rate versus time elapsed after a hypothetical brief burst of star formation. When convolved with the SAD maps described in the previous section, the DTD provides the current SN rate in each region of the galaxy \citep{MaozMan2012, Maoz2014} in the following way:

\begin{equation}
\label{eq:r_i}
R_i = \sum_{j=1}^{N} M_{ij} \Psi_j
\end{equation}

where M$_{ij}$ is the stellar mass formed in cell $i$ in the age-interval $j$ given by the SAD map, and $\Psi_j$ is the DTD value in the age bin $j$. We use the form of the core-collapse DTD given Eq A.2 of \cite{Zapartas17}, which accounts for the effects of binary stellar interactions at $Z_{\odot}$. For Type Ia SNe, we assume the parametric form of the Type Ia DTD from Maoz et al (2012) based on the compilation of all observational constraints to date,

\begin{equation}
\label{eq:psiIa}
\Psi_{Ia}(t) = (4 \times 10^{-13} \mathrm{\ SN\ yr^{-1}\ M_{\odot}^{-1}}) \left(\frac{t}{1 \ \mathrm{Gyr}}\right)^{-1}
\end{equation}

where $t$ is the delay-time between star-formation and SN. The form of the Type Ia and core-collapse SN DTDs are shown in Figure \ref{fig:dtd}. The volumetric SN rate in cell $i$ ($\dot{n}_{SN, i}$ for use in Eq.\ \eqref{eq:sigma})
can then be estimated from R$_i$ as

\begin{equation}
\label{eq:dotsn}
\dot{n}_{SN, i} = \frac{R_{i} \mathrm{cos}(i)}{2 A_{i} z_{sn}}
\end{equation}

where $A_i$ is the cell size of each SAD region ($\approx 83^{\prime\prime} \times83^{\prime\prime}$ or $310 \times 310$ pc$^2$) and $z_{sn}$ is scale height of the vertical distribution of SNe. The factor $\mathrm{cos}(i)$ accounts for the extended line of sight through the disk as a result of the galaxy's inclination angle $i = 77\degree$ \citep{Corbelli2010}, so $z_{sn} \rightarrow z_{sn}/\mathrm{cos}(i)$. We assume $z_{sn}=150$ pc for core-collapse SNe and $z_{sn}=600$ pc for Type Ia SNe, as explained in Section \ref{sec:scaleheights}.

\subsection{Measurements of ISM density and velocity dispersion} \label{sec:measurementofismdensity}
Most of the ISM mass in star-forming regions is in the atomic (\hi) and molecular phases, so we use maps of the 21 cm line of \hi \citep{Braun2009} and the 115 GHz line  $^{12}$CO(J=1--0) \citep{Nieten2006} in M31. The data cubes of \cite{Braun2009} were obtained using the Westerbork Synthesis Radio Telescope (WSRT) and the Green Bank Telescope (GBT), with a spatial resolution of 30\arcsec (or 113 pc at the distance of M31). The \hi column density ($N_{HI}$) 
and non-thermal velocity dispersion ($\sigma_{HI}$) were measured by \cite{Braun2009} from the 21 cm emission along each line of sight assuming a model of an isothermal, turbulence-broadened line profile. We note that evidence for opacity-corrected \hi features in 21 cm is somewhat inconclusive in more recent observations in M31 and M33 \citep{Koch2018, Koch2021}, so we use the opacity-uncorrected map of \cite{Braun2009} (their Fig. 15). The difference in the predicted velocity dispersion from the two different version of the density maps is about $\approx 12\%$, which does not affect our conclusions later. Molecular hydrogen column densities were obtained from the $^{12}$CO(J=1--0) emission  map of \cite{Nieten2006} using the single-dish IRAM 30m telescope. The survey covered $2^{\degree} \times 0.5^{\degree}$ of the M31 disk, yielding a map of CO-line intensity at a final angular resolution of 23$^{\prime\prime}$ (spatial resolution $\approx 87$ pc at the distance of M31). 

The CO-line intensities were converted into H$_2$ column densities ($N_{H2}$)
using the conversion factor $\mathrm{X_{CO}} = 1.9 \times 10^{20}$ mol cm$^{-2}$ (K km s$^{-1}$)$^{-1}$ assumed by \cite{Nieten2006}. The total mass of H$_2$ in M31 is about 14$\%$ that of \hi in the M31 ring. For convenience, we use the \hi velocity dispersion as a proxy for H$_2$ velocity dispersion using the radius-independent ratio of $\sigma_{HI}/\sigma_{H2} = 1.4$ measured by the HERACLES CO and THINGS \hi surveys of nearby galaxies \citep{Mogotsi2016}, as well as in M33 \citep{Koch2019}. We note here that our inclusion of H$_2$ measurements is done to account for the velocity dispersion of the `mass-weighted' ISM, in order to be consistent with the M15 and M16 models, which also predicts the mass-weighted turbulent velocity dispersion.

We combine the density and velocity dispersion of the atomic and molecular phases into an effective mass-weighted ISM. The total mass-weighted non-thermal velocity dispersion in the \hi and molecular phases is then  $\sigma_{obs} = \sqrt{(N_{HI}/N_{tot})\sigma_{HI}^2 + (N_{H2}/N_{tot})\sigma_{H2}^2}$, where $N_{tot}=N_{HI}+N_{H2}$.

We assume the vertical distribution of ISM in M31 is centered on the disk midplane, and approximately Gaussian for the molecular phase and exponential for \hi, consistent with observations of our Galaxy \citep{Dickey1990, Ferriere2001}. Each phase is characterized by an `effective' scale height, which we discuss further in Section \ref{sec:scaleheights}.


For each SAD cell with \hi column density N$_{HI}$, the \hi density along the line of sight $z$ is,

\begin{equation} \label{eq:nh1}
    n_{HI}(z) = \frac{N_{HI}\, \mathrm{cos}(i)}{2 z_{HI}}\mathrm{exp}\left(-
    \left|\frac{z}{z_{HI}}\right|\right)
\end{equation}

As in Eq.\ \ref{eq:dotsn}, the scale height has been corrected for the inclination of M31 with the factor $\mathrm{cos}(i)$. Similarly for each SAD cell with H$_2$ column density N$_{H2}$, the corresponding H$_2$ volume density is,

\begin{equation} \label{eq:nh2}
    n_{H2}(z) = \frac{N_{H2}\, \mathrm{cos}(i)}{\sqrt{2\pi z_{H2}^2}} \mathrm{exp}\left(-\frac{z^2}{2
    z_{H2}^2}\right)
\end{equation}

For ease of interpretation (given our simplified ISM model), we will compare the observed velocity dispersions with the \emph{minimum} velocity dispersion predicted by models. We enforce this by assuming all SNe explode at the mid-plane density, i.e. $n_h = n_h(z=0)$. This is approximately a lower-limit on $\sigma_p$ (we call this $\sigma^{min}_p$) per SAD cell since SNe exploding away from the midplane but still within the scale height of gas would interact with lower densities than at the midplane, deposit greater momenta (Eq.\ \ref{eq:pfin_nh}), and contribute to an effectively higher $\sigma_p$ per SAD cell. This lower limit is also a good assumption since we neglect all other sources of stellar feedback (e.g. winds, cosmic rays) that could add to the momentum budget per SAD cell depending on the environment. Comparing this minimum feedback from SNe with observations leads to some interesting insight as we show in Section \ref{sec:disc}. For each value of ($N_{HI}$, $N_{H2}$), we derive ($n_{HI}$, $n_{H2}$) using Eq \eqref{eq:nh1} and \eqref{eq:nh2}, convert to a total hydrogen mass density $\rho = m_p (n_{HI} + 2n_{H2})/X_H$ in units of g/cm$^{3}$ (where $m_p = 1.67 \times 10^{-24}$ g and $X_H = 0.76$ is the mass fraction of hydrogen), and feed it into Eq.\ \eqref{eq:sigma}. We also take the total number density of hydrogen, in units of atoms cm$^{-3}$ as $n_h = n_{HI} + 2 n_{H2}$, for use in Eq \ref{eq:pfin_nh}.

\subsection{Galactocentric radii of SAD cells} \label{subsec:ring}

We first calculate the deprojected distance of each cell from the center of M31 using the method in \cite{Hakobyan2009}. Let $\left(\alpha,\delta \right)$ be the sky-projected location of each cell centroid, and $\left(\Delta \alpha, \Delta \delta\right)$ be the sky-projected angular offset from M31 center (located at $\alpha_{M31} = 00^h42^m44.3^s$, $\delta_{M31} = +41\degree16^{\prime}9^{\prime\prime}$)\footnote{\url{http://ned.ipac.caltech.edu/}}. Assuming a position angle of M31's disk, $\theta_p$= 38$\degree$ \citep{Corbelli2010}, the location ($u,v$) of each SAD cell in M31's coordinate system is

\begin{align*}
	u &= \Delta \alpha\ \mathrm{sin} \theta_p + \Delta \delta\ \mathrm{cos} \theta_p \\
	v &= \Delta \alpha\ \mathrm{cos} \theta_p - \Delta \delta\ \mathrm{sin} \theta_p
\end{align*}

The radial distance of each cell in the plane of M31 from the M31 center, corrected for M31's inclination ($i = 77\degree$; \citealt{Corbelli2010}), is, 

\begin{equation}
	d^2 = u^2 + \left(\frac{v}{\mathrm{cos}\, i}\right)^2
\end{equation}

where $d$ is the angular distance from the center in arcseconds. We identify the ``ring'' as SAD cells with 10-13 kpc, as shown by the shaded region in Figure \ref{fig:maps}.


\subsection{Assumptions about SN and ISM scale heights} \label{sec:scaleheights}
In this section, we describe plausible ranges and fiducial values for our free parameters: the atomic and molecular scale heights ($z_{HI}$ and $z_{H2}$) and SN scale heights $z_{sn}$ (here on, we will specify the separate scale heights of SNe Ia and CC as $z_{Ia}$ and $z_{cc}$ respectively).

Core-collapse SNe generally occur at lower effective scale heights than SNe Ia \citep{Hakobyan2017}.

\begin{figure*}
\subfigure[]{\includegraphics[width=0.5\textwidth]{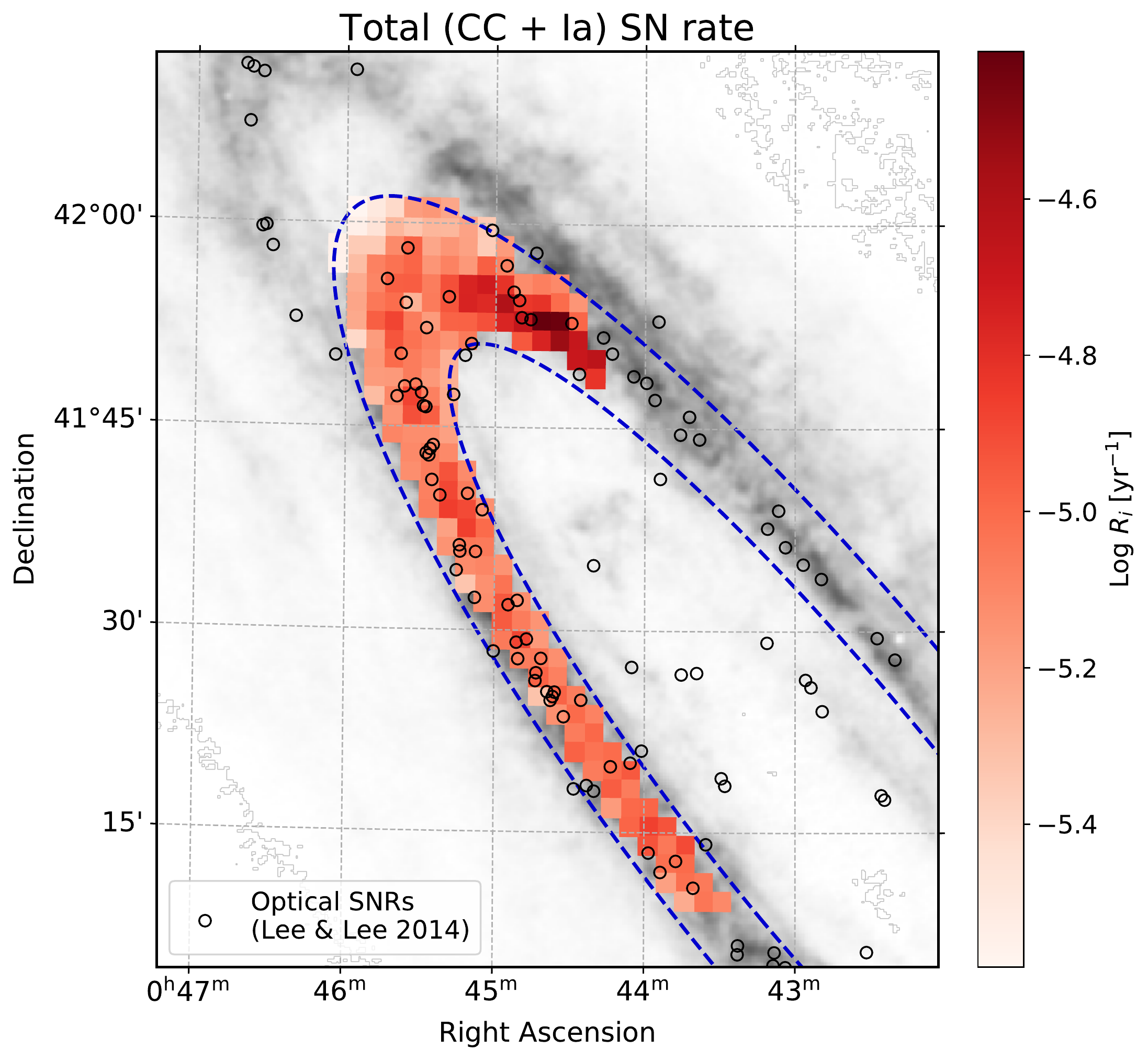}\label{fig:snrate}}
\subfigure[]{\includegraphics[width=0.49\textwidth]{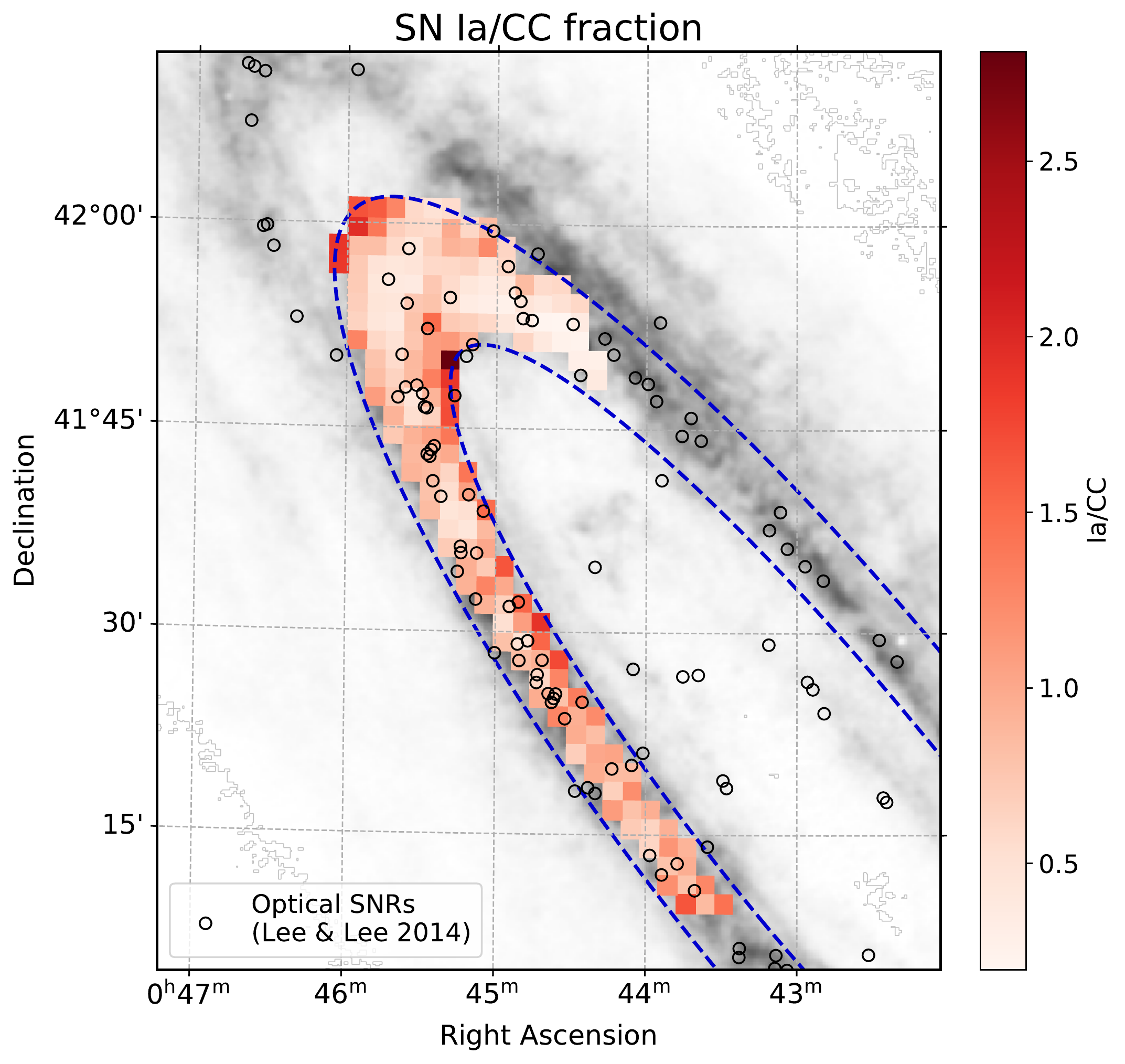}\label{fig:snfrac}}
\caption{(a) The distribution of SN rate (log $R_i$; see Eq.\ \ref{eq:r_i}) in the portion of M31's 10 kpc ring covered by the PHAT survey. Greyscale is the \hi column density map of \cite{Braun2009} (see Section \ref{sec:ratesmodel} for details). The red shading denotes total SN rate (Type Ia + core-collapse) in units of SNe yr$^{-1}$. Black circles show locations of optically-selected SNRs from the \cite{Lee2014} survey, and the blue dashed elliptical contours show the region of the 10--13 kpc ring demarcated for analysis in this paper (see Section \ref{subsec:ring}). (b) The fraction of Type Ia to core-collapse SNe. The figure is the same as in panel (a), except red shading denotes this ratio.}
\end{figure*}

In the Milky Way, open clusters younger than 100 Myrs are all situated within 200 pc of the midplane, with an effective scale height of 60--80 pc \citep{Joshi2016, Soubiran2018}. Since their age and velocity distribution nicely follows that of field stars \citep{Baumgardt2013, Soubiran2018}, we can assume that the general population of core-collapse SN progenitors in the Milky Way also has a scale height of 60--80 pc. However, the disk of M31 is kinematically hotter and more extended than the Milky Way \citep{Ivezic2008, Collins2011}. Based on the ratio of scale heights to scale lengths observed in edge-on disk galaxies \citep{Yoachim2006, Yoachim2008}, \cite{Collins2011} proposed that the M31 disk could be 2-3 times thicker than the Milky Way (although this may be an over-estimate as the galaxies in the \citealt{Yoachim2006} sample are different and less massive than M31). We therefore assume that in M31, 60 pc $<$ $z_{cc}$ $< 200$ pc is a plausible range for the scale height of core-collapse SNe. 

Older ($\sim$Gyr) stars are mostly concentrated in the thin and thick disk with measured scale heights in the range of 140-300 pc and 500-1100 pc respectively in the Milky Way \citep{Li2018, Mateu2018}. The thin disk is slightly younger, with ages in the range of $7-9$ Gyrs compared to the thick disk's age of $\sim 10$ Gyrs \citep{Kilic2017}. The measured shape of the Type Ia SN DTD suggests that progenitors younger than 10 Gyrs will produce the majority of Type Ia SNe \citep{Maoz2010}, so we assume Type Ia SN progenitors are roughly distributed at the same scale height as the thin disk, $\sim$300 pc. This is also consistent with the scale height of SDSS white dwarfs \citep{deGennaro2008, Kepler2017, Gentile2019} and about 4 times the scale height of young core-collapse progenitors, so for simplicity we assume that in M31, $z_{Ia} \approx 4 z_{cc}$, and $z_{cc}$ is in the range mentioned previously.

\hi scale heights in M31 were measured by \cite{Braun1991} in the range of $z_{HI} = 275-470$ pc between radii of 10--13 kpc in M31. We are not aware of any scale height measurements of the molecular phase in M31, but the Milky Way can provide some supplementary information. Studies of the H$_2$ profiles traced by CO in the Milky Way have measured a half-width at half-maximum scale height of 50--80 pc (consistent with being a bit smaller than $z_{cc}$), which is about a factor of 3 lower than the scale height of \hi in the Milky Way \citep{Marasco2017}.

Given these constraints, we can assume that $z_{cc}$ is always less than $z_{Ia}$, $z_{H2}$ is always less than $z_{HI}$, $z_{Ia} \gtrsim 4 z_{cc}$ and $z_{H2} \approx z_{HI}/3$ . Given the range of values allowed by observations, we first analyze our results for a fiducial model where $z_{cc} = 150$ pc, and $z_{HI}=350$ pc, giving $z_{Ia} = 600$ pc, and $z_{H2}=117$ pc. We then change the values of these parameters and their ratios within the plausible ranges discussed previously to assess the impact of assumptions in Section \ref{sec:modelvsobsveldisp}.

\section{Results} \label{sec:results}
In this section, we show the distribution of SN rates across the M31 ring as measured from our SAD map and DTDs, and a comparison of our predicted velocity dispersions predicted by these rates with the observed values along the ring.
\subsection{Distribution of SN Rates} \label{subsec:snrates}

Figure \ref{fig:snrate} shows our SN rate distribution 
estimated from the DTDs and SADs as described in in Section \ref{sec:ratesmodel} (Eq \ref{eq:r_i}). The integrated SN rate in the region we identify as M31's 10 kpc ring is $1.74 \times 10^{-3}$ \snyr, with roughly 39$\%$ contribution from SN Ia and $61\%$ from core-collapse SNe. 

The fraction of this SN rate of Type Ia versus core-collapse is shown in Figure \ref{fig:snfrac}. About 75$\%$ of the ring has a higher core-collapse rate than Type Ia. These regions coincide well with young star-forming regions identified in UV and IR images of M31 \citep{Lewis2017}, and are mostly concentrated in the inner parts of the ring. Regions with the highest core-collapse rates, exceeding that of Type Ia by more than a factor of 3, coincide with the well-known star-forming region OB54, with 
nearly $3.8 \times 10^{5}$ M$_{\odot}$ of stars younger than 300 Myrs \citep[][also see Figure \ref{fig:overpredictedcells} in this paper]{Johnson2016}. SNe Ia generally dominate the total SN rate near the edges of our ring region, coinciding with the inter-arm region as seen in Figure \ref{fig:snfrac}, and exceeding the core-collapse rate by up to a factor of 3 in some SAD cells. 

As evidence of the high characteristic SN rate of the ring, we also show the distribution of optically-selected SNRs in M31 by \cite{Lee2014} in Figure \ref{fig:snfrac}. The majority of SNRs are concentrated along the M31 ring, and particularly associated with regions of higher core-collapse fraction. A more quantitative test of whether the observed SNR distribution is consistent with the SN rates will be the subject of a future paper, since it requires a more rigorous analysis of the poorly understood completeness of SNR catalogs (particularly at optical wavelengths).

\subsection{Comparison of model and observed velocity dispersion}
\label{sec:modelvsobsveldisp}
\begin{figure}
    \centering
    \includegraphics[width=\columnwidth]{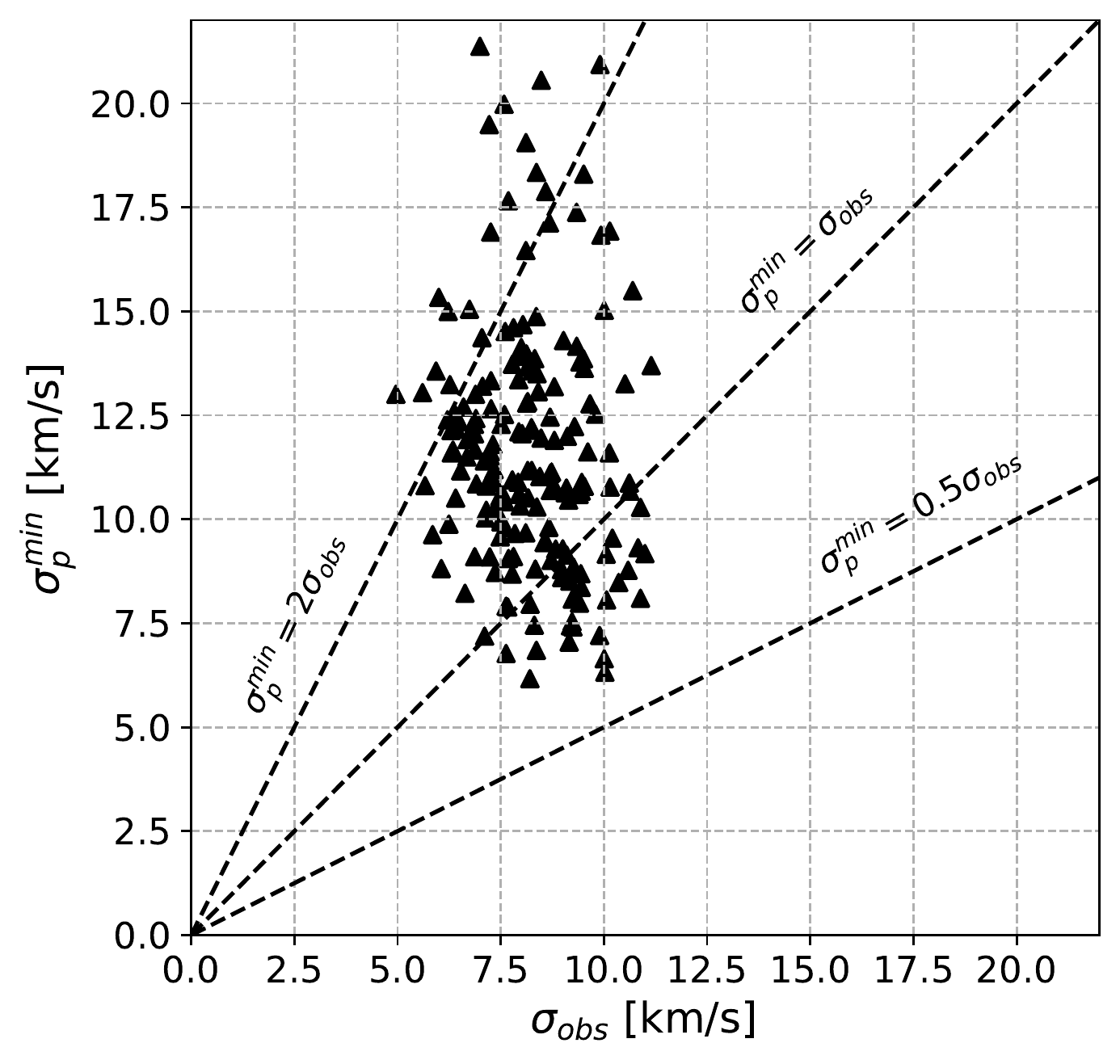}
    \caption{The observed atomic+molecular velocity dispersion in the M31 ring (Section \ref{sec:measurementofismdensity}) compared with our predictions of the minimum velocity dispersion from the fiducial SN momentum-driven turbulence model in Section \ref{sec:predsigma}. The dashed lines indicate values where $\sigma^{min}_p$ is twice, equal and half the $\sigma_{obs}$ values.}
    \label{fig:modelvsobsvel}
\end{figure}
\begin{figure*}
    \centering
    \subfigure[]{\includegraphics[width=0.5\textwidth]{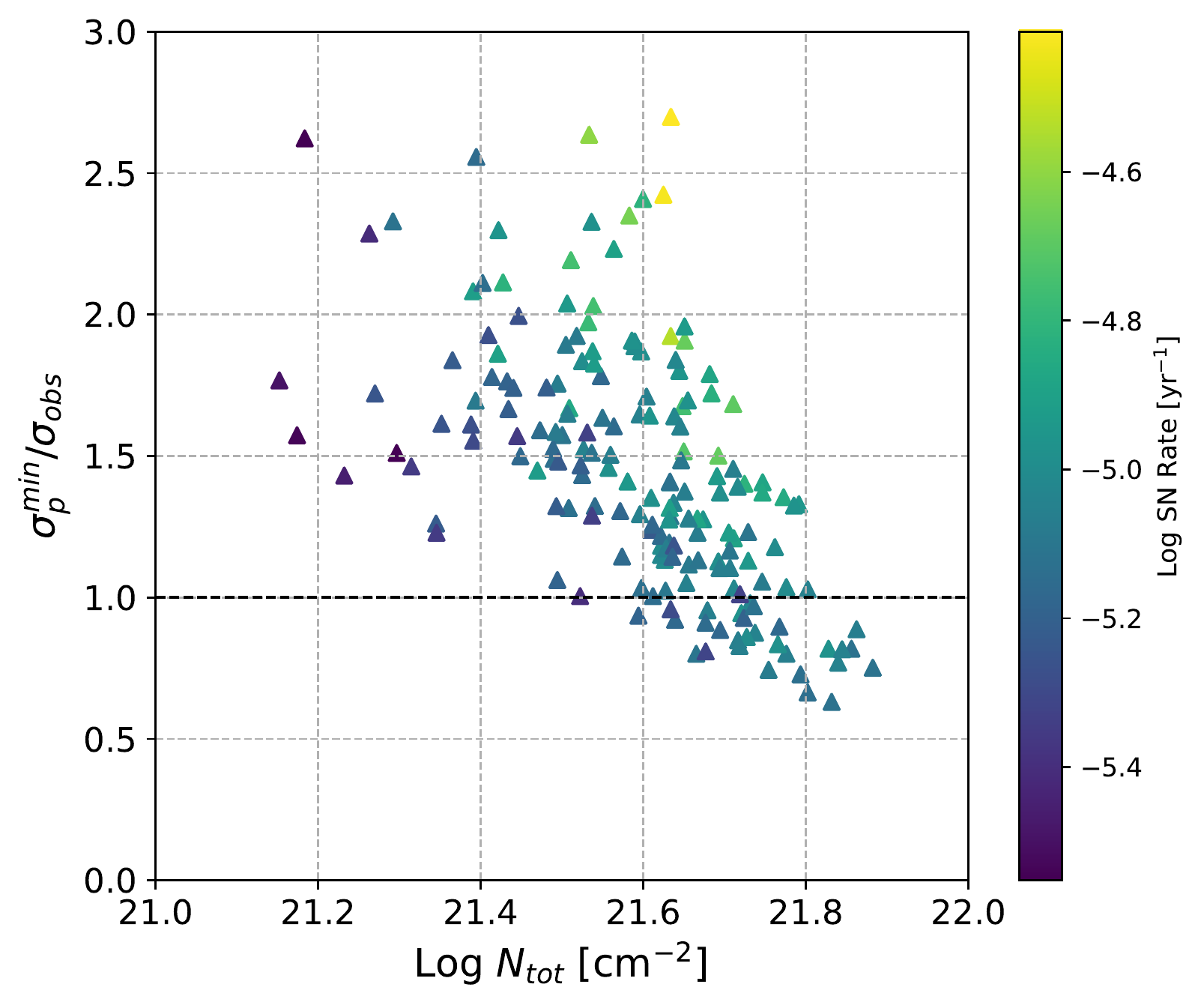}}
    \subfigure[]{\hspace{-0.2cm}\includegraphics[width=0.5\textwidth]{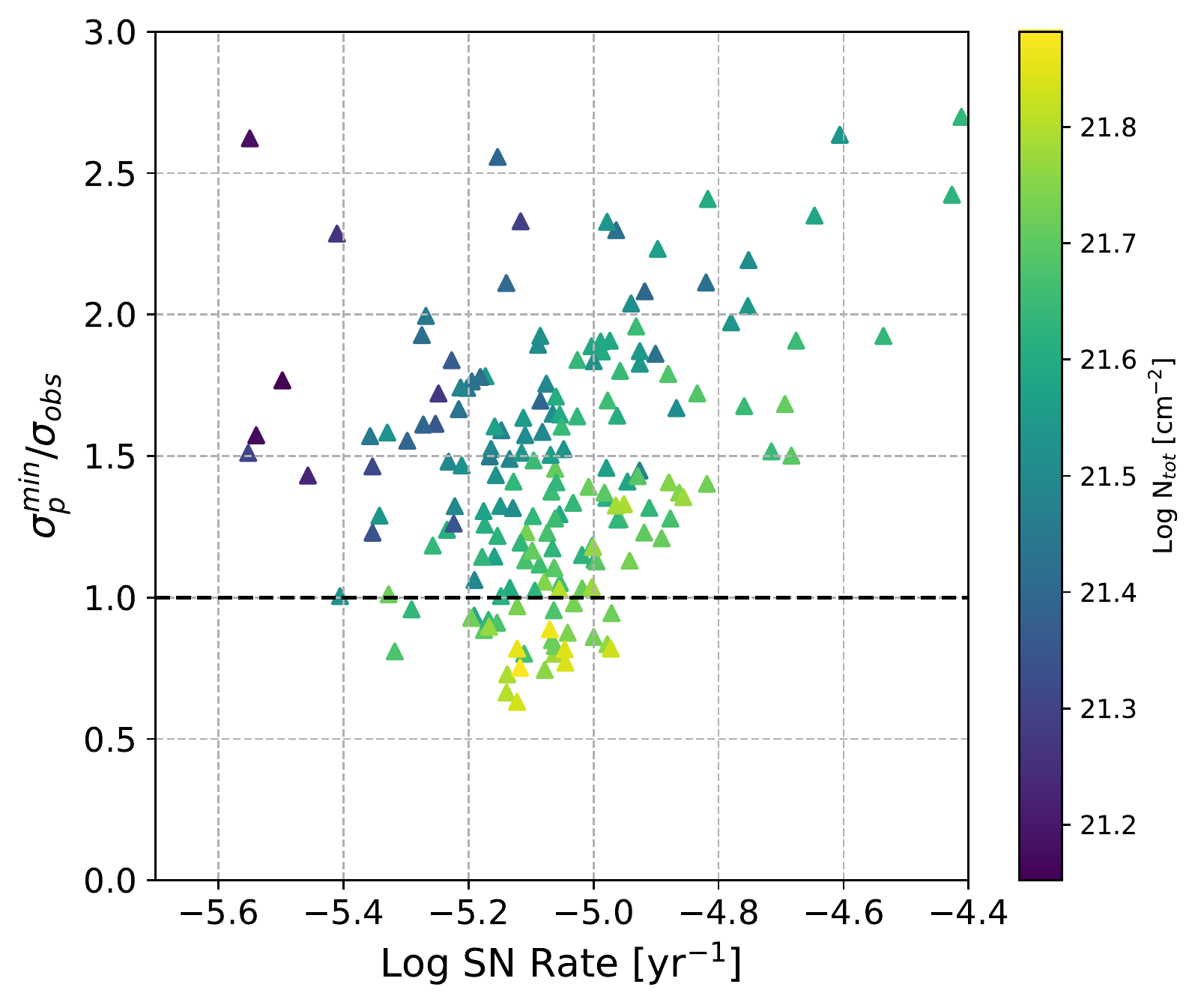}}
    \caption{The ratio of velocity dispersion predicted by our model ($\sigma_{p}^{min}$) and to that measured from 21 cm + CO-line observations ($\sigma_{obs}$) in each SAD cell in the M31 ring (see Figure \ref{fig:maps}) is compared with the observed column density (atomic + molecular; panel a) and the estimated SN rate (panel b) in those cells. Both panels provide the same information, but N$_H$ and SN Rate switches between the x-axis and the colorbar. The points represent the lower limit to the ratio as it corresponds to SNe exploding in M31's midplane (see Section \ref{sec:scaleheights}).}
    \label{fig:sigrationh}
\end{figure*}

We compare the observed ($\sigma_{obs}$) versus mininum predicted velocity dispersion ($\sigma^{min}_p$) in the mass-weighted ISM in Figure \ref{fig:modelvsobsvel}. The observed velocity dispersion exhibits a range of values spanning 4-12 km/s, whereas the predicted values extend up to 20 km/s or higher. On average, we find that for our fiducial model described in Section \ref{sec:scaleheights}, the $\sigma^{min}_p$ mostly exceed the observed values $\sigma_{obs}$, but within a factor of two for 84$\%$ of the SAD pixels in the ring. To understand why our velocity dispersion model over-estimates the observed values in Figure 4., we checked the ratio of observed to predicted velocity dispersion values, i.e. \sigrat against the column density and SN rate, the two fundamental parameters in our model, in Figure \ref{fig:sigrationh}. We find hint of a negative correlation in \sigrat with $N_H$ and a positive correlation with SN rate. 
In particular, SAD pixels with log $(N_{tot}/\mathrm{cm}^{-2})<21.3$ and SN rate or Log $(R_i/\mathrm{yr}^{-1})>-4.7$ mostly have \sigrat$>1$. We examine this more closely in Section \ref{sec:disc}.

\begin{figure}
    \hspace{-0.3in}
    \includegraphics[width=1.2\columnwidth]{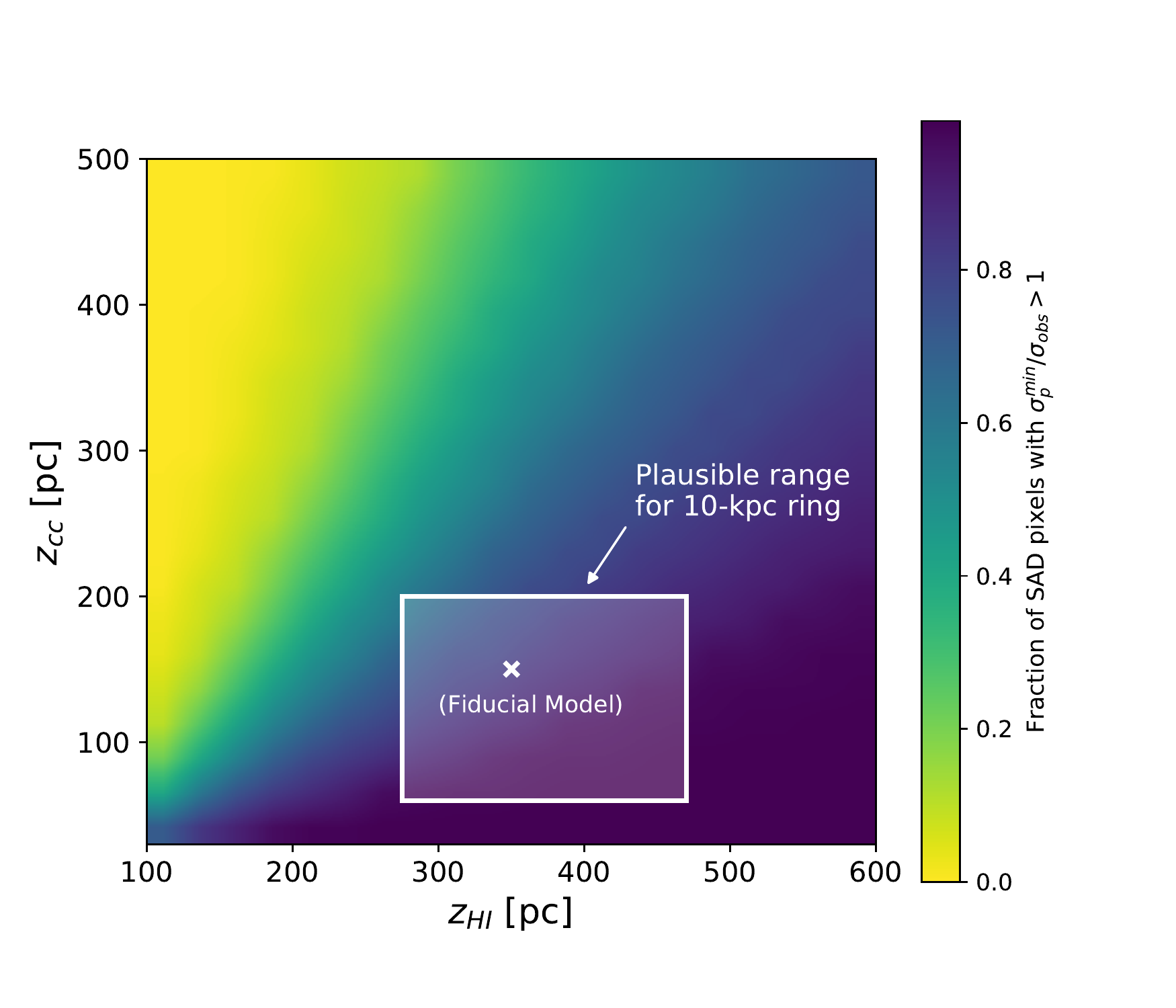}
    \caption{Effect of scale heights on the velocity dispersion calculated from our model. Here, we assume $z_{H2} =  z_{HI}/3$ and $z_{Ia} = 4 z_{cc}$. The plausible values for $z_{cc}$ and $z_{HI}$ in M31, as laid out in Section \ref{sec:scaleheights}, are outlined by the white box. The fiducial model used for predicting values of velocity dispersion in Figure \ref{fig:sigrationh} are shown as a white cross. Colorbar indicates the fraction of SAD pixels with over-predicted mass-weighted velocity dispersion.}
    \label{fig:checkz}
\end{figure}

We briefly discuss the impact of model uncertainties which are certain to alter the \sigrat measurements. The SN rates can vary by an average of 15$\%$ (maximum of 50$\%$), depending on the isochrone model used for constructing SADs \citep{Williams2017}, but this has a relatively small impact on our result. For example, using the MIST SAD solutions (which we have been using), we have about 82$\%$ SAD pixels with \sigrat$>1$, whereas using the PARSEC SAD solutions results in 74$\%$ of SAD cells having \sigrat $>1$, and the correlations with density and SN rates remain. Assumptions about the scale height of gas and SNe directly affect the midplane densities and SN rate densities, which has a larger effect on \sigrat measurements. We therefore assess the impact of varying scale heights on the \sigrat values as as shown in Figure \ref{fig:checkz}. For smaller gas scale heights and larger core-collapse scale heights, 
\sigrat decreases. This is because smaller gas scale heights imply a higher volume density of ISM for a given column density (Eq.\ \ref{eq:nh1}, \ref{eq:nh2}), which reduces the momentum deposition and turbulence driving based on Eq.\ \ref{eq:pfin_nh}. For larger 
SN scale heights, 
the SN rate per unit volume is smaller (Eq.\ \ref{eq:dotsn}), which likewise reduces the momentum deposition rate (Eq.\ \ref{eq:pfin_nh}). Within the plausible range of scale heights  discussed in Section \ref{sec:scaleheights} and marked as a box in Figure \ref{fig:checkz}, about $60\%$ of SAD pixels still have over-predicted velocity dispersion. The part of the parameter space in Figure \ref{fig:checkz} where the fraction of over-predicted cells are below $10-20\%$ involves $z_{cc}>z_{HI}$, which is unlikely given the close association of core-collapse SNe with gas-rich star-forming regions.

\section{DISCUSSION} \label{sec:disc}
\begin{figure}
    \includegraphics[width=\columnwidth]{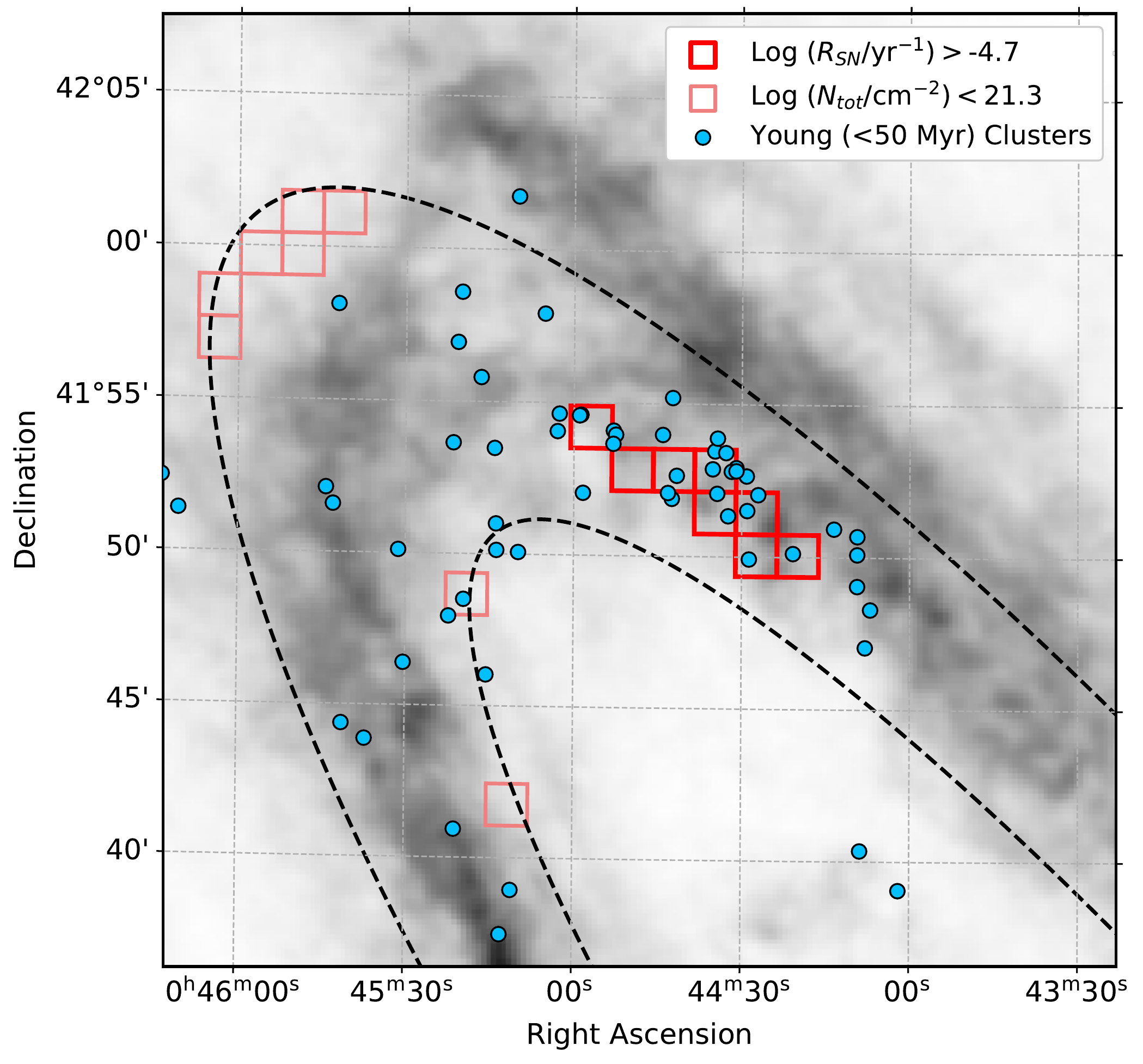}
    \caption{Zoomed-in section of the M31 ring (demarcated by black dashed ellipses) on the grayscale \cite{Braun2009} \hi map, showing 
    the low density (light red) and high SN rate cells (bold red) where the ratio of predicted-to-observed velocity dispersion is mainly above 1 (See Section \ref{sec:disc} for details). Blue circles show locations of star clusters younger than 50 Myr and more massive than $10^3$ M$_{\odot}$ from \cite{Johnson2016}.}
    \label{fig:overpredictedcells}
\end{figure}
\subsection{Insight Into Momentum feedback efficiency}
Our analysis has shown that simple models of ISM turbulence driven by isolated non-overlapping SNRs are consistent within a factor of 2 of observations for most of the star-forming/ISM environment of the M31 ring. Some of the discrepancy can be explained by variation in model parameters (e.g. scale heights of stars and gas) as explained in Section \ref{sec:modelvsobsveldisp}, but in this discussion, we particularly focus on cells with Log $(N_{tot}/\mathrm{cm}^{-2})<21.3$ and Log $(R_i/\mathrm{yr}^{-1})>-4.7$, since all the cells in this range have \sigrat$\gtrsim 1$ even after accounting for plausible variation in SN rates and scale heights in Section \ref{sec:modelvsobsveldisp}.
Regions with \sigrat$>1$ values are interesting in the sense that there are multiple sources of stellar feedback such as stellar winds, radiation pressure, photoionization and cosmic rays, but here the hydrodynamical momentum from SN blast-waves alone over-predict the observed ISM turbulence. 

One reason behind these over-predicted cells could be that $\sigma_{obs}$ were underestimated in our maps, but this is unlikely. More recent, sensitive VLA-based \hi surveys \citep[e.g.][]{Koch2018, Koch2021} suggest that a clean separation of thermal and non-thermal components of the 21 cm line is non-trivial. It is likely that the assumption of \cite{Braun2009} of an isothermal \hi component along the line of sight results in some residual thermal contribution to the non-thermal velocity dispersion. Thus the non-thermal velocity dispersion in M31 we are using from \cite{Braun2009} may be an upper-limit to the actual turbulence contribution.

The regions with \sigrat$>1$, especially in the cells with Log $(N_{tot}/\mathrm{cm}^{-2})<21.3$ and Log $(R_i/\mathrm{yr}^{-1})>-4.7$, may therefore indicate a drawback in the SN momentum-driven turbulence model, so we investigate these regions visually in Figure \ref{fig:overpredictedcells}.
From here on, we couch the column density and SN rate cutoff in terms of a volume density and SN rate surface density cutoff, to be consistent with values used in simulations. The low column density cutoff corresponds to $n_h < 0.2$ cm$^{-3}$ for our fiducial scale heights, and the SN rate cutoff corresponds to a SN rate surface density $\Sigma_{SN}>2.1\times10^{-4}$ \snyrperA. 

The low-density cells are situated at the upper and lower edges of the star-forming ring as shown in Figure \ref{fig:overpredictedcells}. Comparison with Figure \ref{fig:snfrac} shows that these regions also have a higher rate of SN Ia than CC, with an average SN Ia/CC ratio $\approx 1.67$ in these cells. M16 noted that at low densities, SNRs have longer cooling timescales, and may come into pressure equilibrium with the ISM before cooling or depositing a significant amount of momentum \citep{MO77}. This missing physics in our model is likely the reason for the over-predicted $\sigma^{min}_p$. 

SAD cells with Log $(R_i/\mathrm{yr}^{-1})>-4.7$ are spatially correlated with the prominent star-forming region OB54 as mentioned in Section \ref{subsec:snrates}. One possibility, as raised by M15 and M16, is that in high star-forming regions, overlapping shocks from close-proximity SNRs might cancel some of the outgoing momentum (parameterized by $f$ in Eq \ref{eq:sigma}, which we had set to 1). For example, a reduction of more than a factor of 2 in  $\sigma^{min}_p$ is achieved with $f<0.2$, which is consistent, though slightly less than $f=0.3-0.4$ assumed in the SN-driven turbulent ISM simulations of M16. Other possibilities include a non-negligible fraction of the cold ISM mass is driven out by clustered SNe driving a hot, over-pressurized outflow \citep{Sharma2014, Gentry2017}. This explanation is plausible given the detection of X-ray emission in this region by \cite{Kavanagh2020}, and was also given as an explanation by previous energy-balance studies that similarly observed an excess of SN energy over the measured ISM turbulent energy in the central high star-forming regions of galaxies \citep{Tamburro2009, Stilp2013, Koch2018, utomo2019}. A remaining possibility is that a non-negligible ISM mass in these regions is sustained at warmer phases, invisible to 21 cm or CO-line maps, due to the cumulative heating by SNe and pre-SN processes like winds and photoionization.

The results above indicate that fiducial models of momentum feedback from SNe used by most cosmological simulations, which generally assume non-overlapping, non-clustered SNRs, may require adjustment at low densities and at high SN rates due to aforementioned non-linear effects of clustering and SNR evolution at low-densities. This can be quantified by a suppression factor $f(n_h,\, \Sigma_{SN})<1$ for $n_h \lesssim 0.2$ cm$^{-3}$ and $\Sigma_{SN}>2.1\times10^{-4}$ \snyrperA, although a more precise form of this relation will be explored in a subsequent paper where we account for the energy and momentum carried away by any high-velocity outflows or warm diffuse gas from these regions. 

The result also highlights the role of Type Ia SN feedback in low-density regions of the ISM, where it can exceed core-collapse SN rates \citep{Li2020a, Li2020b}. The energetics of Type Ia SNe are particularly pronounced in the central few kpc of M31 (though not explored in this paper), where it is likely responsible for the bright X-ray halo emission and depleted metal abundances in the region \citep{Tang2009, Telford2018}.

\subsection{Comparison with previous studies of ISM energy balance}
As the molecular ISM in our data is only $\sim 14\%$ of the atomic ISM, our results primarily explore feedback in the atomic ISM, and therefore it is interesting to compare our work with previous studies of energy balance in the ISM traced by atomic hydrogen. \cite{Tamburro2009} showed that SN energy alone can drive turbulence in atomic gas within the optical radius of nearby galaxies, with an approximate coupling efficiency of $\sim 10\%$ \citep{Thornton1998, MacLow2004}. Similar results were also obtained by \cite{Stilp2013} with globally-averaged \hi observations. More recently, \cite{Koch2018} and \cite{utomo2019} extended these techniques to M33, with the latter study allowing coupling efficiency to vary with radius.

A key difference between our work and previous ones is that we examine spatially-resolved \hi line profiles along different lines of sight, as opposed to globally or radially-stacked \hi profiles. This allows us to compare the observed ISM turbulence with the local properties of the star-forming and ISM environment.

Phenomenologically, there are a few key differences between our work and previous studies also worth mentioning. Similar to \cite{utomo2019}, we account for turbulence driven by momentum-conserving phase of isolated SNRs, and constrain the efficiency of this momentum-feedback driving the observed turbulence, as opposed to previous studies that considered the efficiency of initial SN energy (= $10^{51}$ ergs) going into the ISM turbulence (the majority of which will be radiated away without impacting the gas). The M15 and M16 models also assume that turbulence is driven at the radius where SNR dissolves into the ISM, which strongly depends on the ISM density (i.e. when $v_s \sim \sigma$). This is different from the assumption of constant driving scale (equal to the scale height) or decay timescale in \cite{Tamburro2009} and \cite{Koch2018}. These assumptions affect the predicted SN feedback. For example, \cite{utomo2019} showed that a spatially-varying decay timescale allowed SNe to drive turbulence in M33 out to 7 kpc instead of 4 kpc, by which point the star-formation rate and gas densities decrease by an order of magnitude compared to the central region. \cite{Bacchini2020} similarly showed that a variable decay timescale makes SNe efficient enough to drive turbulence in the THINGS galaxies throughout, as opposed to just within the optical radius \citep{Tamburro2009}. 

Despite the differences in methodology, our work agrees with previous studies that SN energy driving is inefficient, particularly in regions characterized by high star-formation rates. The higher spatial resolution offered by a more nearby galaxy like M31 reveals that regions where our models disagree with observations also correlate with regions of clustered star-formation, signifying the importance of taking into account clustering effects in SN feedback models. A direct comparison with previous studies is complicated given the differences in methodology, but our work highlights the importance of spatially-resolved observations in Local Group galaxies in the study of SN feedback.

\section{Conclusion}
In this paper, we have tested the paradigm of SN momentum-driven ISM turbulence developed by recent high-resolution vertical disk simulations. We compare model prescriptions with resolved observations of stellar populations and ISM in the prominent 10-kpc star-forming of M31, where stellar feedback is expected to be the main source of turbulence. The spatially-resolved PHAT stellar photometry in the northern third of the disk provides detailed stellar-age distributions (SADs) in $\approx 310$ pc$^2$ cells, which we convolved with known forms of the SN delay-time distribution to predict the core-collapse and Type Ia rates across M31's ring. We used ISM densities of the neutral atomic gas (traced by 21 cm \hi line maps) and molecular gas (traced by $^{12}$CO(1-0) line maps) alongside the SN rates to predict the steady-state mass-weighted turbulent velocity dispersion, using the feedback prescriptions of \cite{Martizzi2015} and \cite{Martizzi2016}. We compared these model estimates against the scaled turbulent velocity dispersion obtained from \hi and CO maps of M31. We assumed all SNe explode in the galaxy midplane where the line-of-sight density is highest, effectively providing a lower-limit on the predicted velocity dispersion. We summarize the following key results from our work :-

\begin{enumerate}
\item We find an integrated rate of $\approx 1.7 \times 10^{-3}$ SN yr$^{-1}$ in the ring covered by PHAT, with 61$\%$ contribution from core-collapse SNe. Regions with dominant core-collapse contribution coincide with known star-forming regions as expected, while regions with dominant Type Ia contribution fall near the edges of the ring.

\item We found that the minimum predicted velocity dispersion exceed observed values in 84$\%$ of the ring covered by PHAT for our fiducial model within a factor of 2. Some of the discrepancy can be explained by varying the assumptions regarding SADs and ISM/SN scale heights within plausible limits, but for densities $\lesssim 0.2$ cm$^{-3}$ and SN rates $>2.1 \times 10^{-4}$ \snyrperA, the discrepancy appears to increase.
\item SAD cells with SN rates $>2.1 \times 10^{-4}$ \snyrperA where velocity dispersion is over-predicted are spatially correlated with dense concentration of young clusters embedded in a bright thermal X-ray region. This supports the possibility of clustering of SNe in this regime, which is not captured in our momentum feedback model. Clustering of SNe can lower the momentum deposited per SN and mass-weighted turbulence in the ISM as a result of converging shocks from adjacent explosions, mass-loaded outflows, or higher mass fraction in warmer ISM phases due to cumulative action of stellar winds and SNe.

\item The low density ($\lesssim 0.2$ cm$^{-3}$) regions where velocity dispersion is over-predicted coincide with the edges of our ring region where Type Ia SNe dominate the injection rate by nearly a factor of 2. However given the overall low SN rate in these regions, it is likely that the discrepancy could be due to isolated SNRs coming into pressure equilibrium with the ISM before significant amount of cooling and momentum deposition takes place---another effect not included in our models.
\end{enumerate}
Our results provide observational support for including adjustments in fiducial subgrid models of momentum feedback, to account for SNR evolution in clustered and in low-density environments. The work underscores the importance of resolved stellar photometry and cloud-scale atomic and molecular ISM observations for assessing feedback models and ISM turbulence. Newer, more sensitive observations at high spectral resolution, such as the VLA maps of \cite{Koch2021} can provide more detailed characterization of turbulence in atomic clouds in M31. Preliminary comparison have shown that the 2nd moment line-widths of the VLA maps are within a factor of 2 of \cite{Braun2009} non-thermal values, although the former do not yet cover the full PHAT area or the M31 ring. We will expand our present analysis in future papers with data from the ongoing Local Group L-Band Survey\footnote{\url{https://www.lglbs.org/home}} which will cover all of M31, as well as M33 and four Local Group dwarfs, providing \hi maps of unprecedented sensitivity at a wide range of spatial resolution. This extension would allow us to test feedback models with spatially resolved observations across a wide range of star-forming conditions, and empirically obtain corrections to the fiducial models to be included in cosmological simulations.

\acknowledgements
S.K.S is grateful to Robert Braun for sharing the WSRT 21 cm maps of \hi density and non-thermal velocity dispersion in M31. SKS and LC are grateful for support from NSF grants AST-1412549, AST-1412980 and AST-1907790. Parts of this research were supported by the Australian Research Council Centre of Excellence for All Sky Astrophysics in 3 Dimensions (ASTRO 3D), through project number CE170100013. We acknowledge support from the Packard Foundation.
E.R-R is supported by the  Heising-Simons Foundation and  the Danish National Research Foundation (DNRF132).

\software{numpy \citep{numpy}, scipy \citep{scipy}, matplotlib \citep{matplotlib}, astropy \citep{astropy1, astropy2}}

\bibliography{Feedback_letter_revised}
\end{document}